\documentclass[12pt]{iopart}
\usepackage{iopams,psfrag,graphicx}

\newcommand{\wdu}[0]{\ensuremath{\omega^{\uparrow}}}
\newcommand{\wud}[0]{\ensuremath{\omega^{\downarrow}}}
\newcommand{\ww}[0]{\ensuremath{\omega}}

\newcommand{\xwu}[0]{\ensuremath{x^{w}_-}}
\newcommand{\xwd}[0]{\ensuremath{x^{w}_+}}

\newcommand{\rhou}[0]{\ensuremath{\rho^{\uparrow}}}
\newcommand{\rhod}[0]{\ensuremath{\rho^{\downarrow}}}

\newcommand{\rhous}[0]{\ensuremath{{\rho^{\uparrow}}^*}}
\newcommand{\rhods}[0]{\ensuremath{{\rho^{\downarrow}}^*}}

\newcommand{\brhon}[0]{\ensuremath{\bar{\rho}}}
\newcommand{\brhou}[0]{\ensuremath{\bar{\rho}^{\uparrow}}}

\newcommand{\trho}[0]{\ensuremath{\tilde{\rho}}}
\newcommand{\trhou}[0]{\ensuremath{\tilde{\rho}^{\uparrow}}}
\newcommand{\trhod}[0]{\ensuremath{\tilde{\rho}^{\downarrow}}}

\newcommand{\vscpu}[0]{\ensuremath{v_{SCP}^{\uparrow}}}
\newcommand{\vscpd}[0]{\ensuremath{v_{SCP}^{\downarrow}}}

\newcommand{\au}[0]{\ensuremath{\alpha^{\uparrow}}}
\newcommand{\ad}[0]{\ensuremath{\alpha^{\downarrow}}}
\newcommand{\bu}[0]{\ensuremath{\beta^{\uparrow}}}

\newcommand{\bd}[0]{\ensuremath{\beta^{\downarrow}}}
\newcommand{\pd}[0]{\ensuremath{p^{\downarrow}}}
\newcommand{\pu}[0]{\ensuremath{p^{\uparrow}}}

\newcommand{\jdu}[0]{\ensuremath{h^{\uparrow}}}
\newcommand{\jud}[0]{\ensuremath{h^{\downarrow}}}
\newcommand{\jnet}[0]{\ensuremath{h^{net}}}
\newcommand{\ju}[0]{\ensuremath{j^{\uparrow}}}

\newcommand{\jd}[0]{\ensuremath{j^{\downarrow}}}

\begin{document}

\title[Shock-dynamics of two-lane driven lattice gases]{Shock-dynamics of two-lane driven lattice gases }
\author{Christoph Schiffmann}
\address{Theoretical Physics Department, Saarland University, D-66123 Saarbr\"ucken, Germany}
\ead{christoph.schiffmann@fu-berlin.de}
\author{C\'ecile Appert-Rolland}
\address{Laboratoire de Physique Th\'eorique, CNRS UMR 8627, Universit\'e Paris-Sud XI, Bat 210, 91405 Orsay Cedex, France}
\ead{Cecile.Appert-Rolland@th.u-psud.fr}
\author{Ludger Santen}
\address{Theoretical Physics Department, Saarland University, D-66123 Saarbr\"ucken, Germany}
\ead{l.santen@mx.uni-saarland.de}

\begin{abstract}
Driven lattice gases as the ASEP are useful tools for the modeling of various stochastic transport processes carried out by self-driven particles, such as molecular motors or vehicles in road traffic.  Often these processes take place in one-dimensional systems offering several tracks to the particles,  and in many cases the particles are able to change track with a given rate.  In this work we consider the case of strong coupling where the hopping rate along the tracks and the exchange rates are of the same order, and show how a phenomenological approach based on a domain wall theory can describe the dynamics of the system. In particular, the domain walls on the different tracks form pairs, whose dynamics dominate the behavior of the system.
\end{abstract}

\pacs{05.60.Cd, 05.70.Ln, 05.70.Fh, 02.50.Ey}

\maketitle

\section{Introduction}

The dynamics of particles on microscopic scales is generically non-deterministic since the motion of the particles is influenced by the surrounding medium. The stochastic motion can either be passive as for Brownian particles or active, i.e. particles are self-driven by processes consuming energy.

In biological systems molecular motors are example of active particles, which combine  directed and stochastic motion \cite{schliwa_w03}. The molecular motors are proteins carrying out a large variety of  intracellular transport processes. Their active dynamics is characterised by a directed stepwise motion along the cytoskeleton, a network of one-dimensional filaments. The energy of this process is provided by the hydrolysis of ATP.
A detailed understanding of the motor dynamics is of obvious biological interest.

Next to the microscopic systems
self-driven particles exist also on larger scales, e.g. models describing
vehicular traffic \cite{helbing01,chowdhury00}. 
The stochastic nature of the motion is in this case not
related to thermal fluctuations, but rather to the integration
of uncontrolled degrees of freedom - the distraction
of a driver, its psychological state, the mechanical
reactions of the car, etc. In the same way, pedestrian or gregarious animal traffic
can be classified in the same family of interacting self-driven particles
described by stochastic processes (see e.g.~\cite{tgf08}).

From a theoretical point of view, it is of interest to extract the
generic non-equilibrium behaviour of these systems.
Systems of interacting self-driven particles show particularly interesting behavior if
open boundary conditions are applied. Then one observes so-called boundary
induced phase transitions, i.e. the boundary particle-reservoirs determine
the bulk-properties of the system \cite{hager01}. The transport properties and particle
distributions are the result of a subtle interplay between boundary
reservoirs and bulk properties of the system, which has been analysed in
great detail. The paradigmatic model for self-driven particles is the
so-called asymmetric exclusion process (ASEP) \cite{krug91,derrida98c}.
In the case of the ASEP particles
perform a directed random-walk on a one-dimensional lattice. Particles
interact via steric exclusion, i.e. lattice sites are either empty or
occupied by a single particle. By using recursion relations the stationary
properties of this model have been exactly calculated. 

Although an exact solution of the model exists for the stationary state, it is instructive to exploit
the picture of competing boundary reservoirs~\cite{kolomeisky98} directly, as it can be extended to non-stationary regimes \cite{santen_a02}: Each particle
reservoir tries to impose a domain of constant density. Then the bulk properties
of the system are determined by the dynamics of the shock, i.e. the position
where one observes the sudden transition from one domain to the other. The
dynamics of the shock can be described as a (biased) random walk with
constant rates, where the left and right hopping rates depend on the flow
realized in the two domains. It is important to notice that the hopping
rates of the shock do not depend on its position, a property which is
closely related to the conservation of particles in the bulk of the system~\cite{parmeggiani_f_f03,parmeggiani_f_f04,evans_j_s03,popkov03,juhasz_s04}.

Variants of the ASEP have been used in order to model a number of important
real empirical systems, including vehicular and intracellular traffic. In
order to apply particle hopping models to vehicular traffic one has to
introduce at least different particle velocities and a finite reaction time
of the drivers via a parallel update scheme \cite{nagel92}. This reaction time
can be enhanced by making the hopping probabilities dependent on
the velocity \cite{appert_s01,barlovic02}. While traffic models based on cellular automata
give today a satisfying description of intra-lane dynamics
\cite{knospe04},
there is a growing interest in the traffic community to
understand better the lateral interactions and the
dynamics of lane-changes.

In the present work we analyse a generalized ASEP which is defined on a
two-lane system. We consider uni-directional motion, i.e. particles
move in the same direction on both lanes.
In the bulk, the particle number is conserved globally,
but this is not the case for each separate lane, since lane-changes are
allowed.  This setup is relevant for vehicular traffic on multilane roads, pedestrian traffic and intracellular transport. 
A model of this kind has been studied recently by Reichenbach et al. \cite{reichenbach_f_f06,reichenbach_f_f07,reichenbach_f_f08} who concentrated on the weak coupling limit,
which is relevant for processive molecular motors.  In this limit the small amplitude of the exchange rates implies a continuous flow profile for each track. 

Here we consider the system for arbitrarily strong lane-changing rates, i.e. exchange rates that are at least of the same order as the hop rates,  which is relevant in case of pedestrian and vehicular traffic, where frequent lane-changes are observed. In this case it is possible to observe not only a discontinuous density profile but also a discontinuous flow profile.
We show here that the system can still be characterized by
a domain wall approach. In particular we concentrate on the
dynamics and synchronisation of the shocks in the system,
and show how they dominate the system behavior.

The paper is organised as follows. First a review of the results obtained in the case of weak coupling will be presented. As preliminaries to the strong coupling case,
a notion of adaptation between the lanes will be defined,
and the possibility to extend the use of second class particles
(SCP) \cite{ferrari92,derrida93a} to two-lane systems will be discussed.

The dynamics of shocks under strong coupling will be
exemplified on a special case, illustrating how shocks
on both lanes are coupled and how their joint motion can be
predicted in a domain wall picture. We also discuss the inner structure of the shock pair
and the relevance of our results for the phenomena taking place in strongly coupled driven systems in general.

\section{A two-lane model}

\subsection{Definition of the model}

\begin{figure}[htbp] 
 \begin{center} 
\psfrag{D}{$D$}
\psfrag{wd}{$\omega_d$}
\psfrag{wa}{$\omega_a$}
\psfrag{p}{$p$}
 	\includegraphics*[scale=0.6]{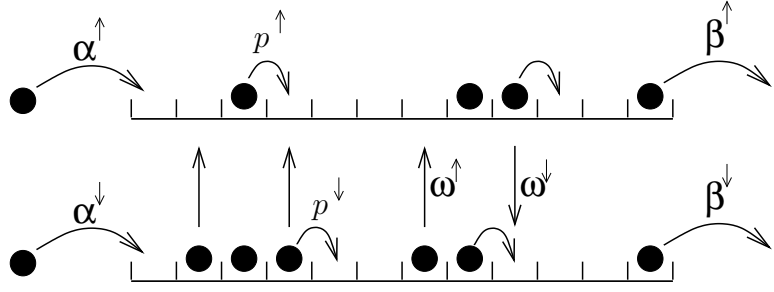} 
   \caption{Schematic illustration of the particle dynamics. }
\end{center}
\end{figure}

In this work we consider a $L \times 2$ lattice with open boundary
conditions. Each lattice site may either be empty or occupied with a 
single particle. In the bulk particles hop  from site $i \to i+1$, where $1\leq 
i \leq (L-1)$, with rate $\pu (\pd)$ in upper (lower) lane if site $i+1$ 
empty. In the remaining of the paper, we shall always use
$\pu=\pd=1$. Particles may also change from the upper to the lower lane with rate
$\wud$ and with rate $\wdu$ in the inverse direction.

The asymmetry between the vertical couplings is defined
by $\wud = \lambda \wdu$. The density of
particles is controlled via left and right particle reservoirs which are
coupled to the chains. Particles enter the system at site $1$ with rate
$\au$  (upper lane) or rate $\ad$ (lower lane), if
the first site is empty. The outflow from the chain
is controlled by the rates $\bu$  (upper lane) or $\bd$ (lower lane).

We shall denote by $\ju$ and $\jd$ the horizontal fluxes
of particles respectively along the upper and lower lane,
while $\jdu$ and $\jud$ stand for the vertical fluxes
from the lower to the upper (resp. upper to lower) lanes
and the net vertical flux is $\jnet \equiv \jdu - \jud$.

\subsection{Mean-field equations}

The model introduced in the previous section is
similar to the one in \cite{reichenbach_f_f07},
and we recall how the mean-field equations can be derived.

The mean-field expressions for these fluxes are
\begin{eqnarray}
\ju(x+\frac{1}{2}) & = &  \rhou(x) \left[ 1-\rhou(x+1) \right] \\
\jd(x+\frac{1}{2}) & = &  \rhod(x) \left[ 1-\rhod(x+1) \right] \\
\jdu(x) & = & \wdu \rhod(x) \left[ 1-\rhou(x) \right] \\
\jud(x) & = & \wud \rhou(x) \left[ 1-\rhod(x) \right] \label{jMF}
\end{eqnarray}

In a {\em stationary} state, mass conservation implies that
\begin{eqnarray}
\ju(x+\frac{1}{2}) - \ju(x-\frac{1}{2}) & = & \jdu(x) - \jud(x)
\label{massc1}
\\
\jd(x+\frac{1}{2}) - \jd(x-\frac{1}{2}) & = & \jud(x) - \jdu(x)
\label{massc}
\end{eqnarray}

We now consider the limit $L \gg 1$ and introduce a new space variable $y = x/L$.  Then the mean-field equations read 
\begin{eqnarray}
\left(1 - 2 \rhou\right) \partial_y \rhou =
\partial_y \ju & = & \wdu L \rhod \left(1-\rhou\right)
- \wud L \rhou \left(1-\rhod\right)
\label{eqrho1}
   \\
\left(1 - 2 \rhod\right) \partial_y \rhod =
\partial_y \jd & = & \wud L \rhou \left(1-\rhod\right)
- \wdu L \rhod \left(1-\rhou\right)
\label{eqrho}
\end{eqnarray}
i.e. we obtained coupled differential equations for the densities. Obviously two independent ASEPs are obtained if the coupling rates $\wud$ and $\wdu$
vanish. 

These equations will be valid only in the
slowly varying regions of the density profile.

\section{Overview of useful notions}

\subsection{Domain wall picture for one-lane ASEP}

The stationary and the transient state of driven lattice gases with open boundary conditions are characterized by a competition between the capacity of the boundary reservoirs and the capacity of the chain. This feature of driven lattice gases has been used in order to develop an effective theory for the macroscopic properties of open systems \cite{kolomeisky98}. This section can be ignored by the reader already familiar with these notions.

The basic idea of the domain-wall-picture is most simply
understood for one-dimensional driven lattice gases with mass
conservation and a single particle-species. The most
prominent example of this type of models is the TASEP
(Totally asymmetric exclusion process) which we will consider
for concreteness. In this case the left reservoir of capacity
$\alpha < \alpha_{max}$ \footnote{The boundary rates are
given as multiples of $p$.  $ \alpha_{max} = 1/2$ for the
ASEP.} establishes a low density domain of density
$\rho_l(\alpha)$ at the left of the system. The current in
the low density domain is given by $j(\alpha)$ (if
$j(\alpha) < j_{max}$) where  $ j_{max}$ is the capacity of
the chain. The low density domain competes with a high
density domain of density $\rho_r(\beta)$ which is controlled
by the right reservoir of capacity $\beta$ if  ($j(\beta) <
j_{max}$). Both domains are separated by a so-called domain
wall or shock which is located at $x_s(t)$ \footnote{Note
that the discussion refers to the continuum limit of the
model. In discrete space we can only distinguish between
empty and occupied sites. }. In general we have 
$\rho_l(\alpha) \neq \rho_r(\beta)$ such that the density profile is discontinuous at
the shock position $x_s(t)$.

In order to characterize  the properties of the system it is crucial to describe the dynamics of the domain wall. The dynamics of the domain wall can be illustrated by considering a half-open chain first, i.e. an ASEP with $(\beta=0)$. We also assume that the chain is initially empty, and thus that
the domain wall is initially positioned at the right boundary. After introducing $N$ particles to the system, the domain wall is shifted $N$-sites to the left assuming that $\alpha \ll p$.
As $N=j({\alpha}) \Delta t$ the average velocity of the domain wall is simply given by $j({\alpha})$.

For arbitrary values of $\alpha,\beta$ one has to consider the displacement for general values of the densities inside the coexisting domains.
In this case $N$ particles correspond to a displacement $\Delta x = N/(\rho_r({\beta})-\rho_l({\alpha}))$ and the average velocity of the domain wall is given by $v_r = j(\alpha)/(\rho_r({\beta})-\rho_l({\alpha}))$.
Analogous the motion of the wall towards the exit has the velocity  $v_l = j(\beta)/(\rho_r({\beta})-\rho_l({\alpha}))$.

 Due to the stochastic nature of the coupling between boundary and chain the motion of the domain wall is well described as a random walk with right and left hopping rates $v_r$ and $v_l$. From this picture, it is possible to derive (with a very good agreement with direct simulations) the density profile, but also more subtle quantities such as the fluctuations of the number of particles in the system \cite{santen_a02} or the largest relaxation time in the system \cite{kolomeisky98,nagy_a_s02}.

This picture is valid for general one dimensional systems with particle conservation in the bulk of the system. 
For systems that are coupled to bulk reservoirs, i.e. for which the particle number is not conserved even in the bulk, the above picture has to be modified. Due to particle exchange the flow inside the domains is no longer a conserved quantity. For such systems the hopping rates of the domain wall are position dependent. This implies for example that the shock can be localized in the bulk at the position $0<x_s<L$ if the bias of the walker changes its sign at $x_s$ \cite{juhasz_s04,evans_j_s03,popkov03}. The present model combines global particle conservation  with particle exchange between the two lanes. Therefore one expects to observe localization effects as well as the coupling between the motion of the two domain walls, one on each lane, in the system. 

\subsection{The case of weak coupling - a review}
Different variants of the present model have been considered recently where
the coupling of the lanes has been realized in different ways. The first
class of models is characterized by hopping rules which respect mass
conservation in {\em each} lane in the bulk
\cite{fouladvand_l99,popkov_p01,popkov_s03}.
In these models the coupling between the two lanes
has been realized via hopping rates depending on the configuration on the
other lane, but there are no particle exchanges between the lanes. 
Models with non-vanishing exchange rates have also been considered before \cite{harris_s05}. Pronina
and Kolomeisky \cite{pronina_k04} studied the stationary state of a two-lane system for
symmetric and asymmetric lane changing rules which are coupled to particle
reservoirs. The in- and output rates are the same for the two chains. As a
consequence shocks are localized at the same position for the two lanes and
the structure of the phase diagram can be obtained by a mean-field analysis. 

More general capacities of the boundary reservoirs have been considered in a series of papers by Reichenbach et al.
\cite{reichenbach_f_f07,reichenbach_f_f08}. In their work the present model has also been discussed in terms of a spin representation. Their focus was on establishing the phase diagram and analysing the systems behaviour for weakly coupled chains, i.e. the case where the exchange rates scale as $1/L$. 
From a technical point of view this assumption implies that 
that the current between the two lanes is also of the order $1/L$. Therefore the flow profiles are continuous at arbitrary positions in the system. This is the main difference to our work since here both the exchange rates as well as the density difference between upper and lower lane are of $\mathcal{O}(1)$.

\section{Adaptation between lanes}

\label{sect_adapt}

Focussing on the domain wall dynamics in the strong coupling limit we choose the model parameters such that no additional boundary effects inside coexisting domains appear, as this will be explained in this section.

For $\alpha < 1/2$ and $\beta < 1/2$,
as we have seen in the previous section,
a one-lane system can be described by the domain wall picture: a domain wall (or density jump)
separates two domains of constant density.
Now the question is: What happens when two flat density profiles on two different lanes 
are coupled ($\ww \neq 0$)?
In the general case, a vertical flux immediately sets in.
Assuming a factorized structure on each lane, the vertical fluxes just after the onset of the coupling are
\begin{eqnarray}
\jdu(x) & = & \wdu \rhod(x) \left[ 1-\rhou(x) \right] \\
\jud(x) & = & \wud \rhou(x) \left[ 1-\rhod(x) \right].
\end{eqnarray}

It may happen that $\rhou$ and $\rhod$ are such that the vertical net flux vanishes.
In this case the densities are conserved on each lane, and the flat profiles are maintained. We define density relations which conserve the densities in both lanes as {\em adapted}.
With a relation $\wud = \lambda \wdu$, the condition for the upper and lower densities to be adapted reads

\begin{equation}
\label{adapt_eq1}
\rhod = \frac{\lambda \rhou}{1+(\lambda-1)\rhou}
\equiv f_{\lambda}(\rhou)
\end{equation}
by definition of $f_{\lambda}$.
We also notice that
\begin{equation}
f_{1/\lambda}(X) = 
f^{-1}_{\lambda}(X).
\end{equation}

In the general case, i.e. if the initial densities are not adapted, a non zero vertical flux occurs for non-vanishing exchange rates. 
In the bulk of the system a new local equilibrium between the lanes is reached
quite rapidly (typically a few time steps 
times the inverse exchange rates
for the parameters of table \ref{paramI}).
The stationary bulk densities are adapted and the vertical net flux vanishes.

At the boundaries, each reservoir tries to impose a density
in the system.
As the upper and lower lane reservoirs are decoupled,
the 
particle densities near the boundaries are not automatically adapted
and it is possible to observe a stationary net flow between the two lanes
in the vicinity of the entrance or exit.  
If the entrance rate on one given lane is $\au$ (or $\ad$), 
it is equivalent to say that the entrance reservoir tries to impose a density $\au$ (resp. $\ad$) in the system. If the densities $\au$ and $\ad$ of the entrance reservoirs fulfill the relation (\ref{adapt_eq1}), then we say that the entrance rates are adapted.
The condition reads 
\begin{equation}
\label{adapt_a}
\ad = \frac{\lambda \au}{1+(\lambda-1)\au}
\equiv f_{\lambda}(\au)
\end{equation}
In the same way, the exit reservoirs try to impose
the densities $1-\bu$ and $1-\bd$ respectively on
the upper and lower lane. Therefore the condition for the exit rates to be adapted is
\begin{equation}
\label{adapt_b0}
1-\bd = \frac{\lambda (1-\bu)}{1+(\lambda-1)(1-\bu)}
\equiv f_{\lambda}(1-\bu)
\end{equation}
or equivalently
\begin{equation}
\label{adapt_b}
\bd = \frac{\bu}{\lambda-(\lambda-1)\bu}
= f_{1/\lambda}(\bu).
\end{equation}
If boundary rates (entrance or exit) are not adapted
 a boundary layer forms where a non zero net flux between
the lanes transfers from  the reservoir densities to the
adapted bulk densities (see an example in
\cite{pronina_k06}).

In the remaining of the paper, we choose only adapted boundary
conditions, in order to avoid these boundary layers, and to focus
on the role of domain walls.
The case of non adapted boundary conditions will be discussed
in section \ref{sect_nonadapt}.
Note that even if entrance and exit boundary rates are
adapted, the adapted densities that the entrance reservoirs
try to impose in the system are not necessarily the same as the 
adapted densities imposed by the exit reservoirs, hence
one observes the formation of domain walls in between.

\begin{table}
\begin{center}
\begin{tabular}{c|c|c|c}
parameter & value & parameter & value  \\
\hline
$\au$ & 1/3 & $\bu$ &  0.176 \\
$\ad$ &  0.2 & $\bd$ &  0.3 \\ 
$\wdu$ &  0.2 & $\wud$ &  $\lambda \wdu$ \\
$\lambda$ &  $\frac{1}{2}$ & & \\
\end{tabular}
\end{center}
\caption{Set (I) of parameters. Boundary conditions are adapted on both sides.}
\label{paramI}
\end{table}

\section{Second class particles in two-lane systems}

In the case of one-lane ASEP systems with mass conservation,
a useful tool for tracking shocks is the introduction of second
class particles (SCP) \cite{derrida_l_s97,boldrighini89}.
A SCP is a passive tracer particle which does not modify the dynamics
of the ordinary (or first class) particles. Its dynamics is defined in  a way that,
in presence of a discontinuity in the density profile of first class particles, i.e. a shock,
the SCP will locate itself at the shock position
(at least, for a certain class of shocks, as this will be explained below).
Thus following the motion of the SCP allows to track the dynamics of the shock.

The problem in using the SCP for this model is due to the fact that mass is not conserved for each lane. Therefore we have to modify the dynamics of the SCP in order to keep its ability to trace shocks also for  two-lane systems.

\subsection{Definition of the dynamics for a second class particle
in a two-lane TASEP}

\begin{figure}[htbp] 
   \centering
 	\includegraphics[width=3in]{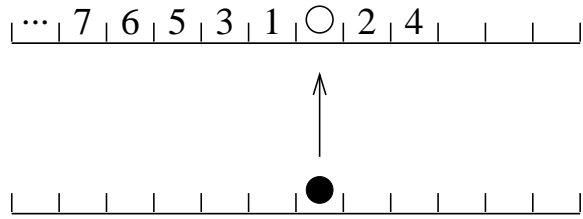} 
   \caption{Schematic illustration of the second class particle (white disk) dynamics,
when a first class particle (black disk) arrives from the
other lane. Neighboring sites are explored in the order indicated by
the figures, until an empty site is found. Then the second
class particle moves to this empty site.}
   \label{fig:scp1}
\end{figure}

The intra lane evolution rules are, as for one-lane systems:
\begin{eqnarray*}
10 & {\longrightarrow} & 01  \\
20 & {\longrightarrow} & 02 \\
12 & {\longrightarrow} & 21
\end{eqnarray*}
where 0, 1, and 2 stand for empty site, first class particle,
and second class particle, respectively. The updates of the SCP are carried out with the hopping rates $\pu$ and $\pd$ of the first class particles. This rule implies that the SCP moves synchronously with the shock \cite{derrida_l_s97} 
 if the number of particles is conserved on each lane.  For the model under consideration this is not the case as (first class) particles are exchanged with the boundary reservoirs and between the lanes. When a first class particle changes lane and the target site is occupied by a SCP one has to make sure that the SCP behaves like a hole. A simple exchange of the positions of first and second class particle is not possible since the SCP cannot change lane.   
Therefore,  non-local moves of the SCP have to be introduced, as the neighboring sites of the SCP are not necessarily empty. 
 In principle several rules are possible. 
  In figure \ref{fig:scp1} we illustrate our choice of  the non-local SCP dynamics. This rule ensures that the SCP does not change the lane and follows the domain wall position in the same lane.

When a first class particle changes lane and arrives on a
site occupied by a SCP, the neighboring sites of the SCP
are explored in the order that is indicated in figure
\ref{fig:scp1}. The SCP has to jump to the first
empty site that is found.

We have chosen an asymmetric move of the SCP in order to avoid the
dynamical trapping of the SCP in the high density region 
to the right of the domain wall. Still, it has to be symmetric enough to avoid a too strong drift towards the left.

\subsection{Mean field calculation of the drift velocity
of a SCP in a constant density background}

In this section, the drift velocity of a SCP placed in a flat
density profile will be calculated 
(the term "density" always refers to a density of
first class particles). We note $\brhon \equiv 1-\rho$.
In a mean-field approximation, the drift velocity of
a SCP located on the upper lane would be
\begin{eqnarray*}
\vscpu &=& (+1)\brhou + (-1)\rhou\\
&+& \wdu \rhod
\left[-\brhou + \rhou \brhou
-2 {\rhou}^2 \brhou
+2 {\rhou}^3 \brhou
-3 {\rhou}^4 \brhou
-4 {\rhou}^5 \brhou
- \cdots \right] \\
&=& 
1 -2\rhou - \wdu \rhod
\frac{\left[1-3\rhou+5{\rhou}^2-7{\rhou}^3+9{\rhou}^4-4{\rhou}^5\right]}
{1-\rhou}
\end{eqnarray*}
If the upper and lower densities are adapted,
using (\ref{adapt_eq1}), $\vscpu$ is a function
only of $\rhou$ and $\wdu$.
In the special case of uncoupled lanes ($\wdu=0$), the drift
velocity $\vscpu$ vanishes exactly for $\rhous = 1/2$.
It is positive for a lower density and negative for a higher one.
Thus the SCP locates itself on shocks that separate
a left low density ($<$1/2) domain and a right high density ($>$1/2)
domain.
Note that shocks that do not go through the density 1/2
cannot be tracked.
When the coupling is non zero, the density $\rhous$ for which
$\vscpu$ vanishes is given by an implicit expression that can
be solved numerically.
The upper SCP will thus locate itself on shocks separating
a left domain with density less than $\rhous$ and a right
domain with density greater than $\rhous$.

In the same way, the density $\rhods$ for which
$\vscpd$ vanishes is defined implicitely by
\begin{equation}
0 = 1 -2\rhod - \lambda \wdu f_{1/\lambda}(\rhod)
\frac{\left[1-3\rhod+5{\rhod}^2-7{\rhod}^3+9{\rhod}^4-4{\rhod}^5\right]}
{1-\rhod}
\end{equation}
and thus the densities at which SCPs localize may be slightly different
on the upper and lower lanes - though this difference turns out
to be negligeable
for the couplings that we are considering here.

In particular, for the set of parameters of table \ref{paramI}
that will be used later, we find that $\rhous = 0.481$
and $\rhods = 0.480$. This is not too far from the one-lane value $0.5$.
Note however that the value of $\rhous$ depends on the coupling strength.
Indeed, this is one reason why we have chosen the coupling constants
of tables \ref{paramI} and \ref{paramII}.
For a too strong coupling, 
the value of $\rhous$ could be too close to the
density of the right domain to ensure properly the
tracking of the shock.

\section{The physics of strong coupling}

In order to explain the main phenomena involved in strongly coupled two-lane systems, we selected one representative set of parameters, given in~\ref{paramI}. The influence of the different model parameters is discussed in the text below.

The influence of the shock is nicely illustrated in the transient regime. We therefore discuss behaviour of the model in terms of the different steps towards the stationary state. 

\subsection{Initializing the system: Decoupled lanes}

We choose an initial condition such that the upper lane is in the high density (controlled by the exit rate) regime and the lower lane in the low density (controlled by the entrance rate) regime. More precisely for the entrance and exit parameters of table \ref{paramI}, and decoupled lanes, the bulk density in the upper lane (resp. lower lane) is given $\rhou_{bulk} = 1-\bu = 0.824$ (resp. $\rhod_{bulk} = \ad = 0.2$).

The parameters are chosen such that the domain wall picture is valid,   i.e. the entrance and exit rates are chosen such that they do not exceed the capacity of the chain. The domain wall is localized towards the entrance in the upper lane, and towards the exit in the lower lane.
In order to track the dynamics of the domain walls we place one SCP on each lane.
The decoupled system including the SCP is simulated until a stationary state is reached.

\subsection{Early stage - local adaptation}
\begin{figure}[htbp] 
\centerline{
(a) \includegraphics[width=3in]{fig3a.eps} 
(d) \includegraphics[width=3in]{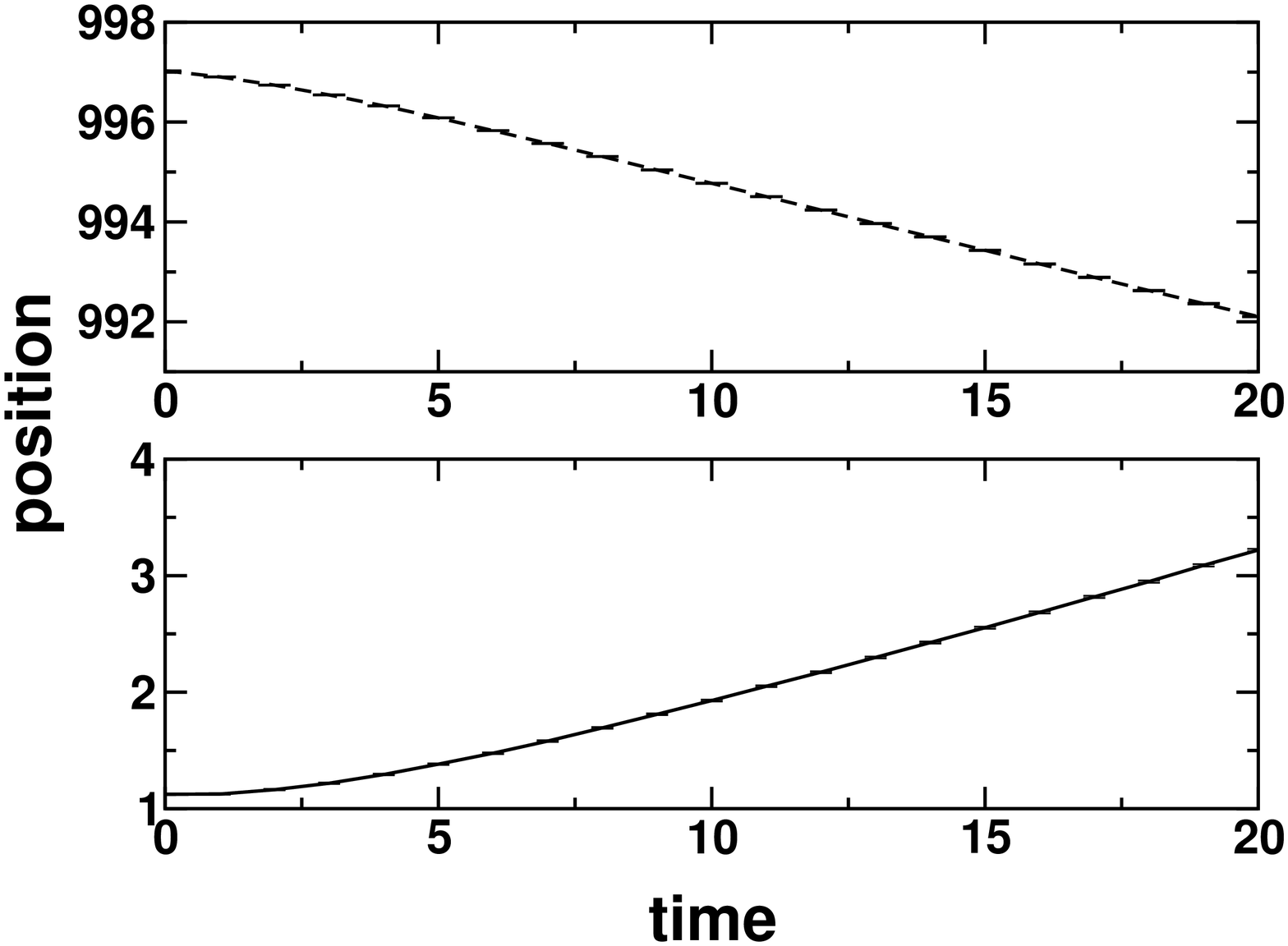} 
}
\vskip 0.5cm
   \centerline{
(b) \includegraphics[width=3in]{fig3b.eps} 
(e) \includegraphics[width=3in]{fig3e.eps} 
   }

\vskip 0.5cm

   \centerline{
(c) \includegraphics[width=3in]{fig3c.eps} 
(f) \includegraphics[width=3in]{fig3f.eps} 
        }
\caption{
Simulation results obtained for the parameters of table \protect{\ref{paramI}} and for a system size $L=1000$.
The left figures (a-c) show the density profiles at different times,
while the right figures (d-f) show the SCP locations in the corresponding time intervals.
For all figures, solid (resp. dashed) lines refer to the upper (resp. lower) lane.
Fig~(a)~: density profiles at time 0 (most upper and lower curves,
black online) and 20 (red online).
Fig~(b)~: density profiles at time 20 (black, red online)
 and time 2304 (grey, green online).
Fig~(c)~: density profiles at time 2304 (grey, green online) and
time 30000 (black, blue online).
In each figure, light grey lines indicate 20 intermediate density profiles.
All data have been averaged over $10^5$ trajectories.
Error bars are given for the mean SCP trajectories; they are smaller
than the line thickness and can hardly be seen.
} \label{fig56} \end{figure}

At $t=0$, some coupling is added between the lanes (see parameters in table \ref{paramI}). As outlined in section \ref{sect_adapt}, this implies that a vertical flux immediately sets in.  
This is illustrated in Figure (\ref{fig56} - a) showing the evolution of the density profile
in the first 20 time steps. In the bulk, particles are exchanged between the lanes
until a local equilibrium is obtained, i.e. until the net flux between the lanes vanishes.
This implies that at the end of this stage, the adaptation relation (\ref{adapt_eq1}) is valid,  i.e $\rhod_* = f_{\lambda}(\rhou_*)$. The actual values of the bulk densities can be obtained from local mass conservation:

\begin{equation}
\rhou_*+\rhod_* = 1-\bu+\ad \equiv \rho_0
\end{equation}
where the r.h.s. is the sum of bulk densities for uncoupled lanes and the l.h.s. is the sum of bulk densities 20 time steps after coupling is activated. 

From these two relations, we find that the bulk densities after 20 time steps should be given by 
\begin{eqnarray}
\rhou_* & = & \frac{\rho_0}{2} - \frac{\lambda+1}{2(\lambda-1)} \pm
\sqrt{\frac{1}{4}\left[\frac{\lambda+1}{\lambda-1}-\rho_0\right]^2+\frac{\rho_0}{\lambda-1}} \\
\rhod_* & = & \rho_0 - \rhou_*
\end{eqnarray}
where the '+' sign is valid if $\lambda > 1$ and  '-' if $\lambda < 1$ (for the special case $\lambda=1$ one gets $\rhou_* = \rhod_* = \rho_0/2$).
In our case, $\lambda=1/2$, and the predicted bulk
density values are $\rhou_* = 0.598$ and $\rhod_* = 0.426$,
which are to be compared with the numerical values
obtained in the simulations $\rhou_* = 0.599\pm0.002$ and $\rhod_* = 0.424\pm0.002$
(see Fig. \ref{fig56}-a).

\subsection{Shock dynamics}

At the end of the local adaptation stage, a new domain wall appears on the right of the upper lane (resp. on the left of the lower lane ), to separate the new bulk domain from the exit driven (resp. entrance driven) domain. There are now four shocks in the system, located at both ends of each lane.

The second class particles allow to track the shocks which are at $t=0$ on the left
of the system in the upper lane and on the right of the system in the lower lane.
The two other shocks cannot be traced by a SCP since they do not go through the density 1/2.

The two shocks on the right (resp. left) of the system are immediately coupled, and start to move in pairs towards the center of the system. Their drift velocity can be computed
analytically (see next subsection).
While the vertical flux is vanishing within each domain, it is nonzero between the two shocks of a given pair. This gives another signature of the motion
of the pair, which can be compared with the SCP tracking.

Eventually both shock pairs merge on each lane (see figures \ref{fig56}-b and \ref{fig56}-e). For our setup the two shock
pairs merge in the bulk of the system around time 2300.
The resulting pair of walls move towards the left boundary since the global capacity of the left reservoirs exceeds the capacity of the right reservoirs. In the final stationary state, it stays localized near the entrance (figures \ref{fig56}-c and \ref{fig56}-f).

In figure \ref{fig56} the time-dependent average position of the SCP is given.
Actually, the SCP trajectories vary from one realization to
another one, since it depends on the realisation of the
noise. The whole time-dependent distribution of SCP-positions
can also be obtained numerically. When the shock is localized
on the left (final state), the width of this distribution
stabilizes to a small value.
The fluctuations of the shock velocities in different realisations of the transients states imply that the merging event does not occur at the same time. Hence quite large variations in the SCP positions from one realization to another one.
Nevertheless, as we shall see, the average SCP motion
can be predicted with high accuracy.

\subsubsection{Mean velocity of a pair of domain walls}

\begin{figure}
\centerline{
\includegraphics[width=0.7\columnwidth]{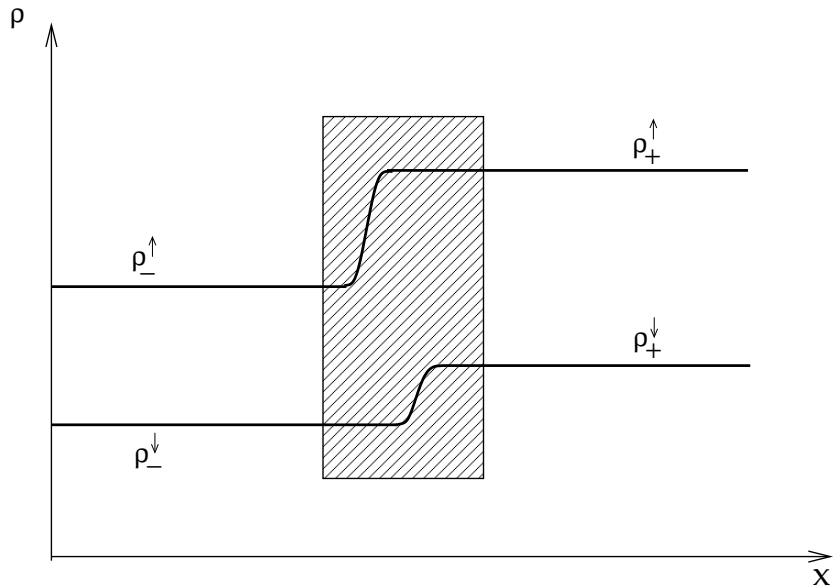}}
\caption{Sketchy representation of the domain wall picture
for two strongly coupled lanes. The walls on both lanes
move together, though they are not necessarily exactly at the
same location. Ignoring the details of the structure
in the vicinity of the shocks, one can draw a black box
around it. On each side, the densities on both lanes are supposed to be
adapted.}
\label{strong_walls}
\end{figure}

The velocity of a pair of walls can be computed by considering
the pair as a whole (see sketch on figure \ref{strong_walls}) neglecting its internal structure. On each side of the "black box" surrounding the pair of shocks,
densities are supposed to be adapted. Factorized states are
again assumed on each side of the walls.

We refer to quantities on the left (resp. right) of the walls
with a superscript `-' (resp. `+'). If we note
\begin{equation}
j^{tot}_{\pm}(x) \equiv \ju_{\pm}(x) + \jd_{\pm}(x)
\end{equation}
the total horizontal flux in the left and right domains, and
\begin{equation}
\rho^{tot}_{\pm}(x) \equiv \rhou_{\pm}(x) + \rhod_{\pm}(x)
\end{equation}
the total densities, then the wall velocity is given
\begin{equation}
V = \frac{j^{tot}_{+} - j^{tot}_{-}}{\rho^{tot}_{+} - \rho^{tot}_{-}}
\label{velocity}
\end{equation}
due to mass conservation.
In case of the left pair of walls, the densities to the left
of the walls are $\au=1/3$ and $\ad=0.2$, and the densities on the
right of the walls are $\rhou_* = 0.598$ and $\rhod_* = 0.426$.
This yields a velocity
\begin{equation}
V = \frac{\rhou_*(1-\rhou_*) + \rhod_*(1-\rhod_*)
- \au (1-\au) - \ad (1-\ad)}{\rhou_*+\rhod_*-\au-\ad} = 0.209
\end{equation}

In the same way, the velocity of the right pair of walls is
$V = -0.260 $. This is in excellent agreement with the measured velocities
of the second class particles, $0.209$ and $-0.261$ respectively
(from figure \ref{fig56}-e).

After the merging of both pairs of walls, the resulting pair of walls
separates a right domain with densities $1-\bu$ and $1-\bd$ respectively on
the upper and lower lane, and a left domain with densities $\au$ and $\ad$ respectively on the upper and lower lane. The predicted drift velocity of the pair
of shocks is now $-0.02745$, and the shocks move to the left until they
reach the boundary.
Again, this velocity can be measured on the simulations
(see fig. \ref{fig56}-f), and we find $-0.02745 \pm 10^{-5}$, in agreement with
the theoretical prediction.

\subsubsection{Structure of the shock region}
In the previous sections, we considered the pair of
walls as a whole around which we can draw a black box.
In this section, we would like to calculate, in the
spirit of domain wall calculations, the internal structure
of the shock region, and in particular the distance between
the shocks.

Outside from the pair of shocks, density profiles are
flat, with adapted densities, so that the net vertical flux is vanishing.
By contrast, in the region between the shocks, density profiles
are x-dependent. The densities on both lanes are not adapted,
and as a result there is a net flux from one lane to the other.

We refer to this region as the "shock region" in the remaining of the paper.
Due to their intra-lane dynamics, shocks would like to separate
but the vertical flux acts as glue to keep them together.
First we consider the case $V=0$, i.e.
the shock pair does not undergo any drift.

\begin{figure}
\centerline{
\includegraphics[width=0.6\columnwidth]{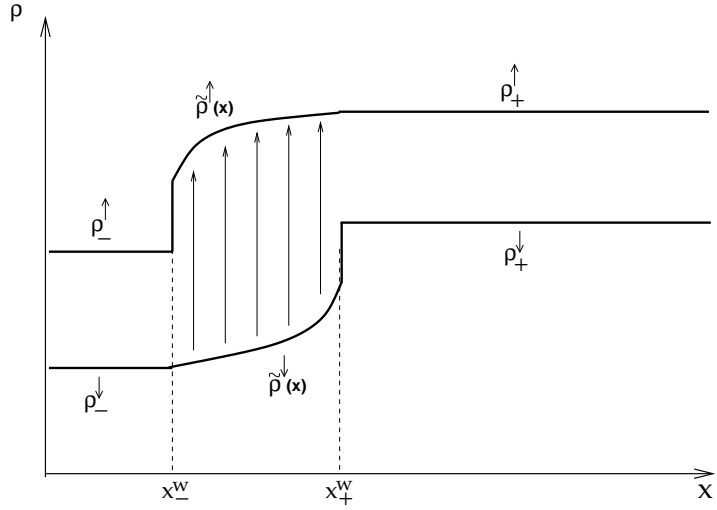}}
\caption{Sketchy representation of the structure of a
pair of walls. We assume that in a first approximation,
we can represent the walls as real discontinuities,
separated by a varying density profile $\trho$.
}
\label{fig_struct}
\end{figure}

Figure \ref{fig_struct} gives a sketch of the
simplest idealized structure that we expect.
In a stationary state, the sum of the horizontal
fluxes must be conserved through the left discontinuity.
As $\jd$ and $\rhod$ are continuous in $\xwu$,
$\ju$ must also be continuous in $\xwu$.

It should be noted that, under the coarse grained description
that was used in the previous section, the horizontal
flux on each lane was discontinuous through the shock pair.
Now that we observe the shocks at a smaller scale,
for which the inner structure of the pair is explicitly
described,
only the density profiles are discontinuous,
while the horizontal flux profiles must be continuous
on each lane. Due to this zooming procedure, we recover
the same level of description
as the one that was used for weak coupling in
\cite{reichenbach_f_f07}, and thus 
the calculations are similar in the case $V=0$.

The only way to obey the continuity constraint
for $\ju$ is to have $\trhou(\xwu) = 1-\rhou_{-}$.
As $1-\rhou_{-}$ and $\rhod_{-}$ are not adapted, 
a vertical flux is initialised in the shock region.
In the same way, we must have 
$\trhod(\xwd) = 1-\rhod_{+}$ on the left of
the right discontinuity. Between $\xwu$ and $\xwd$, the densities $\trhou$
and $\trhod$ must obey the coupled differential equations (\ref{eqrho1},\ref{eqrho}) with the boundary conditions

\begin{eqnarray}
\trhou(\xwu) & = & 1-\rhou_{-}  \\
\trhod(\xwu) & = & \rhod_{-}  \\
\trhou(\xwd) & = & \rhou_{+}  \\
\trhod(\xwd) & = & 1-\rhod_{+} 
\end{eqnarray}

Only two of these boundary conditions are necessary
to calculate the solution numerically starting for example at $x=\xwu$.
Of course, the solution that is obtained depends on
the coupling strength. Matching the other boundary conditions will give the
relation between the coupling strength and the width
of the shock region $\xwd-\xwu$.

If $V\neq0$, the density profiles inside the shock
region are not determined anymore by the stationary
coupled equations (\ref{eqrho1},\ref{eqrho}). Still, we assume that the shock region
has a structure which is preserved during
the drift motion. This validity of this assumption will be discussed later.
The assumption implies that all points
of the structure must be translated with the
same velocity.

Between the shocks, the density 
can be expressed by
\begin{eqnarray}
\rhou(x) & = & \trhou(x-Vt) \\
\rhod(x) & = & \trhod(x-Vt)
\end{eqnarray}
where the new $\trhou$ and $\trhod$
functions have to be determined.
The mass conservation equation equivalent
to (\ref{massc1},\ref{massc}) but including now the time
dependence is

\begin{eqnarray}
\rhou(x-Vdt) - \rhou(x) & = & \left[\ju(x-\frac{1}{2}) - \ju(x+\frac{1}{2}) + \jdu(x) - \jud(x)
\right] dt\\
\rhod(x-Vdt) - \rhod(x) & = & \left[\jd(x-\frac{1}{2}) - \jd(x+\frac{1}{2}) + \jud(x) - \jdu(x)
\right] dt\\
\label{massc2}
\end{eqnarray}

Again, introducing a new space variable $y = x/L$,
we find for the continuous limit when $L$ becomes large

\begin{eqnarray}
\left(-V + 1 - 2 \rhou\right) \partial_y \rhou
 & = & \wdu L \rhod \left(1-\rhou\right)
- \wud L \rhou \left(1-\rhod\right) \\
\left(-V + 1 - 2 \rhod\right) \partial_y \rhod
& = & \wud L \rhou \left(1-\rhod\right)
- \wdu L \rhod \left(1-\rhou\right)
\label{eqrho2}
\end{eqnarray}

The boundary conditions for these coupled differential
equations are obtained from the relations
\begin{eqnarray*}
V & = & \frac{\trhou(\xwu)\left[1-\trhou(\xwu)\right]
- \rhou_- \left[1- \rhou_-\right]}{
\trhou(\xwu)- \rhou_- }  \\
& = & \frac{\rhod_+\left[1-\rhod_+\right]
- \trhod(\xwd)\left[1-\trhod(\xwd)\right]}{
\rhod_+ - \trhod(\xwd) }  \\
& = & \frac{\rhou_+\left[1-\rhou_+\right]
+ \rhod_+\left[1-\rhod_+\right]
- \rhou_-\left[1-\rhou_-\right]
- \rhod_-\left[1-\rhod_-\right]}{
\rhou_++ \rhod_+- \rhou_-- \rhod_-}
\end{eqnarray*}
and from the fact that
$\trhod(\xwu) = \rhod_-$ and $\trhou(\xwd) = \rhou_+$.

Again, we have 4 boundary conditions for two coupled
first order differential equations, and the overdetermination
will allow to determine the width $\xwd-\xwu$ of the shock
region (for a given coupling strength). 
When the coupling rates are of the same order as
the particle hopping - i.e. of order 1 -, then
the shock region $\xwd-\xwu$
is finite and the dynamics of the shocks are
strongly coupled.

This is in contrast to the case of weak coupling,
for which the coupling strength scales as $1/L$, and
the distance $\xwd-\xwu$ is thus of the same
order as the system size.
In the latter case, already discussed in detail by 
Reichenbach et al \cite{reichenbach_f_f08},
the density profile becomes non-uniform
in the bulk.

\subsubsection{Distance between the shocks of a pair}

We have measured the mean distance between the shocks
in the stage corresponding to fig \ref{fig56}-c,
i.e. we address the time interval where only one pair of shocks is present in the system,
which has not yet reached the left boundary.

As the merging event does not occur at the same time
in all realizations, the measurement was done
between $t_1$ and $t_2$, where $t_1$  and $t_2$ are
realization dependent. $t1$ is the time at which
the left SCP reaches its rightmost position,
and $t2$ is the time at which the left SPC is located at
a distance of 50 sites from the left boundary.

We measure that $\langle t_1\rangle = 2371 \pm 1.4$ (with a standard
deviation equal to $420$) and $\langle t_2\rangle = 16333 \pm 8.7$
(standard deviation $2745$). 
The mean distance between the SCPs during
this time interval $t_1 < t < t_2$ is $\langle d \rangle = 4.44 \pm 0.015$.
However, this distance fluctuates : the standard deviation
for d is $4.8$. These fluctuations can be due to two combined effects.
First, the dynamics of the SCPs are such that SCPs perform non local
jumps, which can be quite large. Second, the distance between the shocks of the pair
may also fluctuate in time. To say it otherwise, the variations of the distance between
SCPs can be due to real variations of the distance between
the shocks of the pair, and/or to a lack of precision in the tracking
of the shocks.

In order to distinguish between the two effects,
several types of tracking of the shocks could be compared.
Different dynamics for the SCP can be defined. 
An alternative shock-tracking method  uses the fact that the net vertical flux is not vanishing between the pair of shocks. 
We leave this comparison for future work.

\subsection{Final stationary state}

For the set of parameters of table \ref{paramI},
the final stationary state is characterized by
the localization of the final pair of shocks
near the entrance boundary.
From figure \ref{fig56}-f, we get the final SCP positions,
resp. 4.9 and 7.9 on average on the upper and lower lane.

These values could be compared with the localization lengths
for the decoupled lanes, which are equal to 
 $\xi = \left |\ln \left(\frac{\beta (1-\beta)}{\alpha (1-\alpha)}\right)\right |$.
With the parameters of table \ref{paramI},
we find $0.426$ for the decoupled upper lane (wall near the entrance) and $0.272$ for the decoupled lower lane (wall near the exit).

It means that here, the localisation length of the pair of shocks is much larger
than what it would be for a single shock in the decoupled lane.
This is not surprising, as the shock on the lower lane is pulling
the pair towards the bulk.
Note however that we cannot exclude that, once again, the higher value
that we find for the localization length is an artefact
of the tracking by SCP.
Another way of measuring the localization length would be through
the density profile. An exponential fit, which is in good agreement with the simulation results, leads to a localization length
equal to 7.1 and 7.6 on the upper and lower lane.

In general deviations from the exponential profiles may occur since the applied domain wall picture does not consider the internal structure of the shock. Surprisingly for the present choice of parameters the deviations from the exponential profile are rather small, although the width of the pair of shocks is of the same order of magnitude as the localisation length.

\section{Delocalization of the shocks}
\label{other}
\begin{figure}[htbp] 
   \centerline{ (a) 	\includegraphics[width=3in]{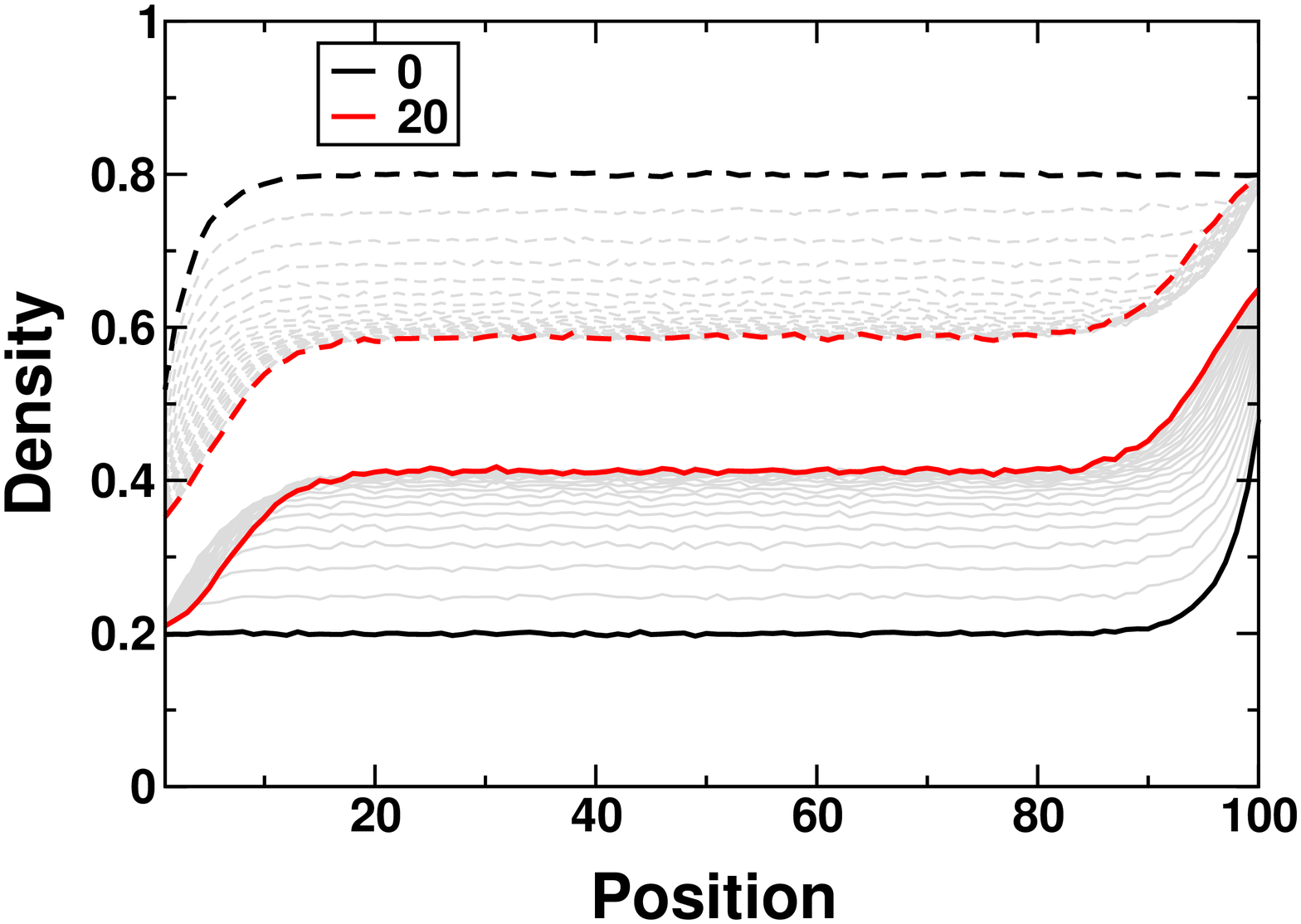}     }
\vskip 0.5cm
   \centerline{(b) 	\includegraphics[width=3in]{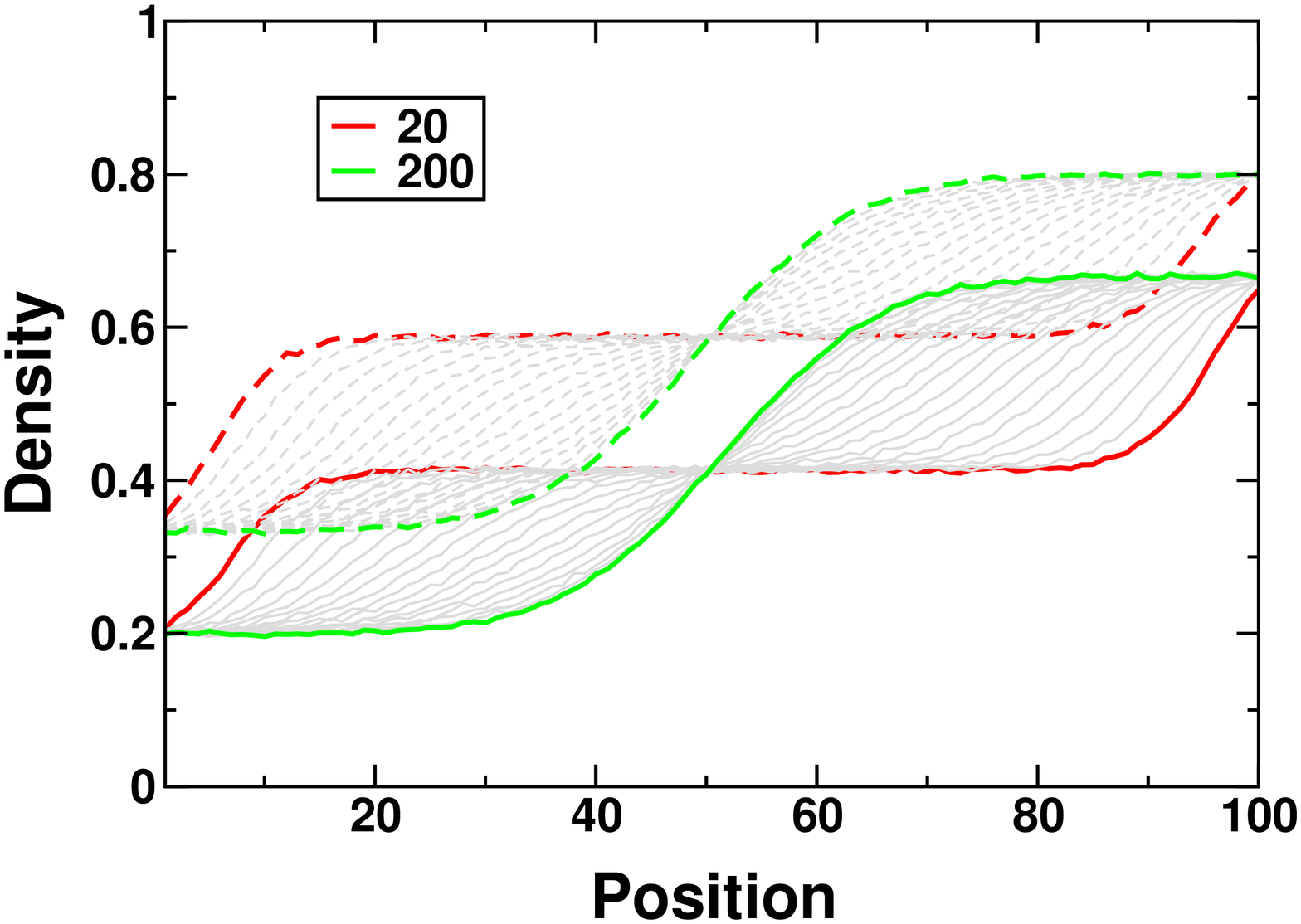}        }
\vskip 0.5cm
  \centerline{ (c) 	\includegraphics[width=3in]{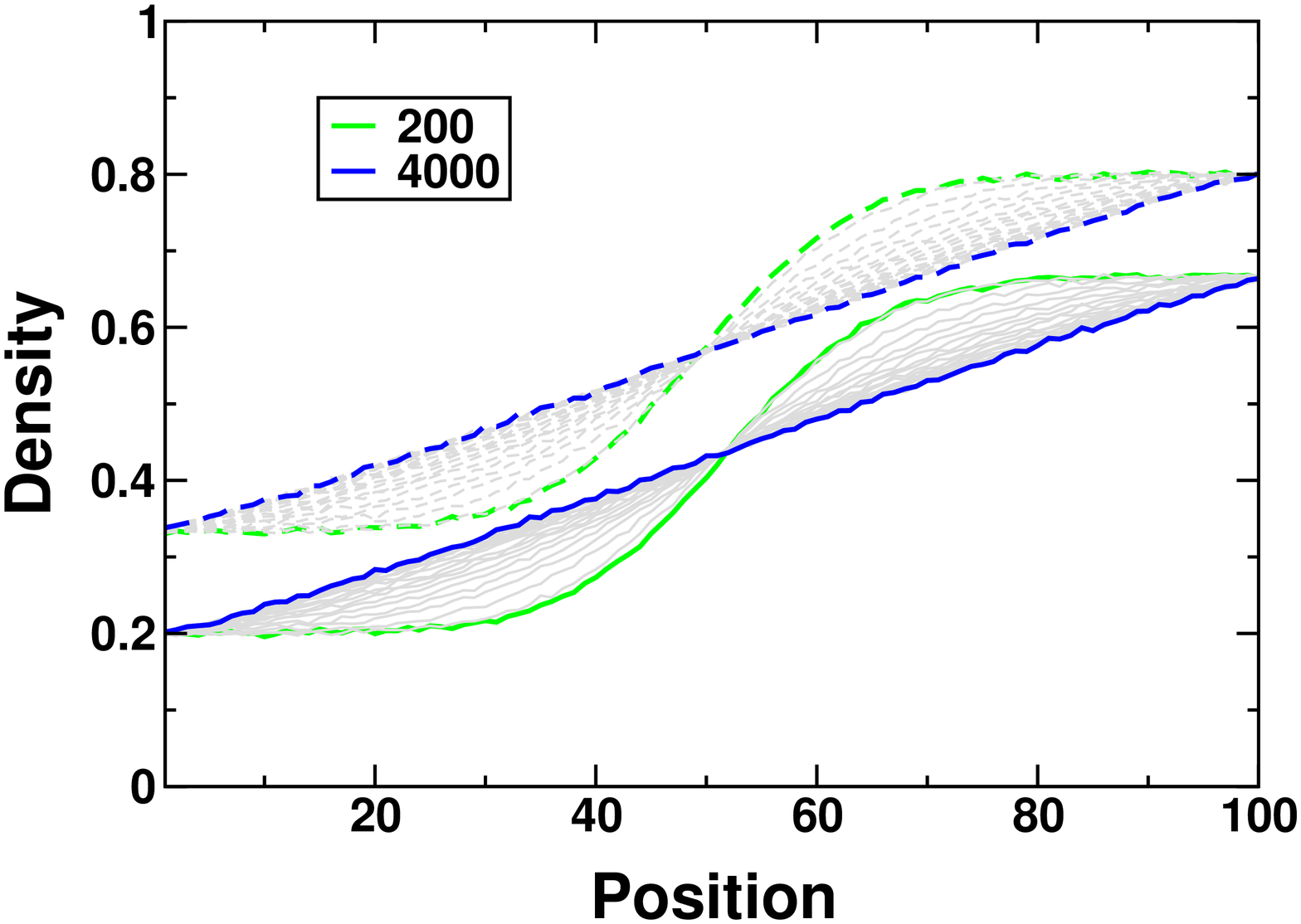}    }
   \caption{Average density profiles at different times,
for the parameters of table \protect{\ref{paramII}} and
for a system size $L=100$. Dashed (resp. solid) lines refer to the upper (resp. lower)
lane.
Fig (a) : density profiles at time 0 (most upper and lower curves,
black online) and 20 (red online).
Fig (b) : density profiles at time 20 (black, red online)
 and time 200 (grey, green online).
Fig (c) : density profiles at time 200 (grey, green online) and
time 4000 (black, blue online).
Thin grey lines give density profiles at some intermediate times.
The final stationary profile is linear and interpolates between the densities
imposed by the reservoirs.}
   \label{fig:linear}
\end{figure}

The set of parameters of table \ref{paramI} is such that shocks
are localized at the left boundary in the final stationary state.
Some other behaviors can be observed, 
for example, in the special case where $\bu = \ad$ and $\bd = \au$,
the drift velocity of the final pair of shocks is exactly zero. 
Thus it undergoes a non biased random motion forever,
and the resulting density profile is linear, as can be seen on fig. \ref{fig:linear}
 for the parameters of table \ref{paramII}.
More precisely, the linear density profile is exactly a linear
interpolation between the densities imposed by the reservoirs,
as this can be predicted by the domain wall picture.

\begin{table}
\begin{center}
\begin{tabular}{c|c|c|c}
parameter & value & parameter & value  \\
\hline
$\au$ & 1/3 & $\bu$ &  0.2 \\
$\ad$ &  0.2 & $\bd$ &  1/3 \\ 
$\wdu$ &  0.2 & $\wud$ &  $\lambda \wdu$ \\
$\lambda$ &  $\frac{1}{2}$ & & \\
\end{tabular}
\end{center}
\caption{Set (II) of parameters. Boundary conditions are adapted on both sides.}
\label{paramII}
\end{table}

\section{Transition from weak to strong coupling}
\label{transition}
\begin{figure}[htbp] 
   \centering
 	\includegraphics[width=4in]{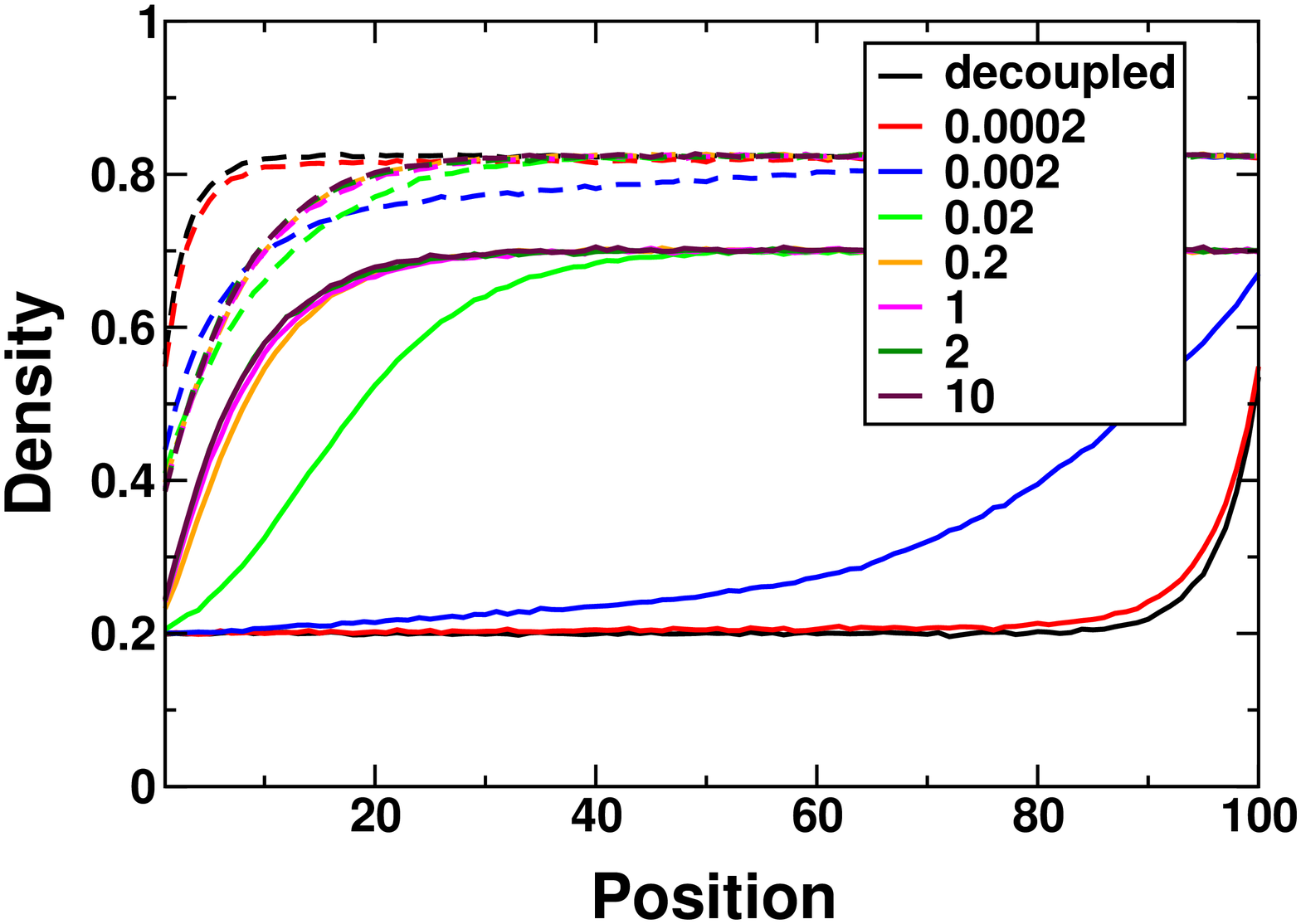}
   \caption{Final stationary density profiles for the parameters
of set (I), except for the parameter $\wdu$ which is varied
from $0$ to $10$. Here the system size is $L=100$.
Solid (dashed) lines correspond to the upper
(lower) lane. }
   \label{fig:trans}
\end{figure}

For the set (I) of parameters that we chose,
the shocks in the decoupled system are located at
opposite ends in the system, while coupling tends
to keep the two shocks together.
If the coupling strength is increased from zero,
the first effect dominates at first and shocks
remain at each end of the system, until the
"glue" between the shocks is strong enough to keep
them together. Figure \ref{fig:trans} illustrates such transition
from weak to strong coupling. In the upper lane, the shock is always localized on the
left of the system. In the lower lane, there is a transition
from a right to a left localization.

\section{Localization and adaptation}

\label{sect_nonadapt}
In the whole paper, we have considered only adapted
boundary conditions. In this section we briefly discuss the physics of the model 
for more general model parameters. 
Non adapted boundary conditions create adaptation layers which
interpolate between the non adapted densities imposed by the reservoirs,
and the bulk adapted densities.

In the case of weak coupling (exchange rates scaling as $1/L$),
adaptation layers
extend over a non vanishing fraction of the system.
By creating such layers on both sides of the system,
it is possible to localize shocks in the bulk.
 This phenomenon is similar to the localization
observed in one-lane ASEP models without mass conservation
\cite{juhasz_s04,popkov03,parmeggiani_f_f03,parmeggiani_f_f04,evans_j_s03}.
Another way to characterize this regime would be to say that
the width of the shocks scales as the system size,
and that the density variations that we observe are
actually the internal structure of a bulk localized shock.
It is not possible anymore to draw a black box around the
shocks, and no shocks dynamics as the one described in this paper can
take place.

\begin{figure}[htbp] 
  \centering
	 \includegraphics[width=5in]{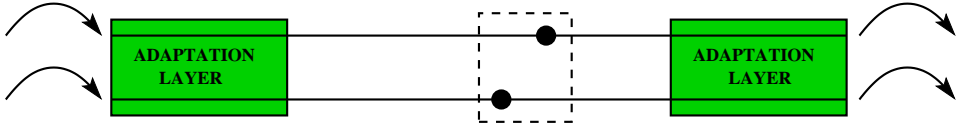} 
   \caption{Schematic illustration of the different regions of the system in the case of strong coupling.
   Grey (green online) areas represent the adaptation regions.
The pair of shocks (represented as dots in the bulk)
is surrounded by a dashed box. }
   \label{fig:scheme}
\end{figure}

In the case of strong coupling, 
the different capacities of the boundary reservoirs lead to an adaptation layer of a given size $l_{al}$, which is determined by the amplitude of the exchange rates, and which is independent of the
system size - at least if $L \gg l_{al}$.
The different regions of the system are illustrated in Fig.  \ref{fig:scheme}.
If the pair of shock is located in the bulk of the system, a black
box can be drawn around it, and the shock velocity
can be determined by (\ref{velocity}) since the densities on both sides of the shocks are adapted (and thus uniform in space).
No localization is possible in this bulk region.
Near the boundaries however, highly non-trivial dynamics is observed. 
If the shocks enter the shaded adaptation layer, the velocity of the shocks
become position dependent.
This implies e.g. that the shock might be unable to reach the boundary sites of the system, instead it would be reflected somewhere inside the adaptation layer. 
Then in the final state, the pair of shocks would not be localized exactly on the boundary, but at a finite distance from the boundary.
Thus localization would also take place, but not in the bulk.
Besides, the width of the pair $\xwd-\xwu$ would not necessarily be small
in front of the adaptation layer width, so the inner structure of the pair
 would also
be modified in the adaptation layer, and the shocks on each lane
could be localized at different positions. We are not able at this stage to
predict quantitatively the final localized state, but we expect
the overall picture to be correct.

Now we could wonder what would happen for exchange rates that scale as $1/L^{\alpha}$ ($\alpha < 1$).
We expect in this case the adaptation layers to have a width increasing with $L$. However, they should occupy a vanishing fraction of the system when
$L$ becomes large.
Thus, some shock dynamics would be possible in the bulk, and localization
would occur only in the adaptation layers on each side of the system.
To say it otherwise, the extension of the shock pair could become very large,
but not enough to prevent some motion in the bulk.  The transition from weak to strong coupling can be investigated more systematically by applying a pertubative approach in $1-\alpha$.

\section{Summary and conclusion}

Stochastic models of self-driven particles describe the behaviour of many transport processes on different scales. The self-driven particles often use one-dimensional structures as tracks, e.g. biopolymers or, at a larger scale, highway lanes.  Although being one-dimensional, often several tracks are offered to the dynamic particles. 

In this work, we consider the case of two-lane systems with particle exchanges between the lanes. We focus on the case of strong coupling, i.e. particle
exchange rates are of the same order as hopping rates.
In contrast to previous work we concentrate on the transient regime rather than the stationary behaviour.
We show that, in the strong coupling limit, the transient regime can be fully understood in terms of 
a domain wall picture, i.e. in terms of the dynamics of pairs of shocks.
 For concreteness we mainly discussed two sets of parameters which are chosen in order to illustrate the generic behaviour of the model in the case of strong coupling.

Our results show  that the transient regime can be divided into two main stages.
The first stage is a very rapid local adaptation between the lanes.
This stage occurs on a time scale which is 
determined by the amplitude of the exchange rates, while the
bulk densities at the end of this stage
are determined by the ratio of the exchange rates.
After this first rapid stage, the exchange flux between the lanes is zero almost
everywhere in the system, i.e. everywhere except in some layers
near the boundaries, as this will be discussed below. 

In the second stage, we have shown that the dynamics is dominated
by the dynamics of several pairs of shocks,
which undergo drift motion,
merge when they meet another pair, and drift again until
the unique final pair reaches the boundary where it stays localized.
In this slow relaxation stage, the pairs of shocks have a dynamics
very similar to the one observed in mass conserving systems.
The non-conservation of mass in each lane is concentrated
between the shocks of a same pair, i.e. the vertical flux
is zero everywhere in the bulk except between the shocks of a same pair,
where it acts as a ``glue'' that keeps the two shocks together.
In this stage, in order to track the shock positions,
we have generalised the dynamics of the
so-called second class particles to two lane systems.
These results underline the flexibility of the domain wall approach for one-dimensional transport.

Near the boundaries, the densities imposed by the reservoirs
can prevent local adaptation, at least in finite size layers
called adaptation layers.
In this paper, we have chosen the parameters of the model such that the boundary reservoirs are adapted, i.e. the net-flux between the lanes is zero also near the system boundaries, and there are no adaptation layers.
This choice allows us to characterise the relaxation of the system purely via the domain wall dynamics.
However, our results could be easily extended to non adapted reservoirs
as, in the case of strong coupling, adaptation layers have a constant width
- and thus represent a vanishing
fraction of the system size when the system size increases.
Then the dynamics of the shock pairs is modified only in the adaptation
layers, but remains unchanged in the bulk.

In the weak coupling limit, adaptation layers invade the whole system,
and bulk localization of the shocks can be observed~\cite{reichenbach_f_f07,reichenbach_f_f08}.
Here, for strong coupling, adaptation layers can prevent the shock pairs from reaching the
boundaries, and can localize the pairs at a fixed distance from the
boundary - but no localization is possible in the bulk.

While our focus was here on the domain wall dynamics in the transient
regime, \cite{tailleur_e_k10} have developped a systematic approach
to determine the phase diagram for the final stationary state,
in the case of adapted boundary conditions,
which should shed more light on the final coupling state between the walls.

In principle, the domain wall approach should allow
to determine the stationary phase diagram in the strong coupling
case, though we chose to address only the non-stationary
dynamics here.
This is a supplementary illustration of the predictive
power of the domain wall picture, even in relatively
complex systems.

\vspace{\baselineskip}

\ack

We thank the DFG Research Training Group GRK 1276 for financial support.

\section*{References}
\bibliographystyle{unsrt}

\bibliography{trafic,neq,motors}

\end{document}